\documentclass[%
 reprint,
 amsmath,amssymb,
 aps,
]{revtex4-1}

\usepackage{graphicx}
\usepackage{dcolumn}
\usepackage{bm}

\begin{document}

\preprint{APS/123-QED}

\title{Sub-$\mu$eV Decoherence-Induced Population Pulsation Resonances in an InGaN system}%

\author{Cameron Nelson$^{1}$, Yong-Ho Ra$^{2}$, Zetian Mi$^{3}$, Paul Berman$^{1}$ and Duncan G. Steel$^{1}$}
\affiliation{%
1. The H.M. Randall Laboratory of Physics, The University of Michigan, 450 Church St., Ann Arbor, MI 48109 \\
2. Department of Electrical Engineering and Computer Science, McGill University, 3840 University St., Montreal, QB H3A 0E9  \\
3. Department of Electrical Engineering and Computer Science, Center for Photonics and Multiscale Nanomaterials, The University of Michigan, 1301 Beal Ave., Ann Arbor, MI 48109
}%

\date{\today}

\begin{abstract}
We report on high frequency resolution coherent nonlinear optical spectroscopy on an ensemble of InGaN disks in GaN nanowires at room temperature. Sub-$\mu$eV resonances in the inhomogeneously broadened third order ($\chi^{(3)}$) absorption spectrum show asymmetric line shapes, where the degree of asymmetry depends on the wavelength of the excitation beams. Theory based on the Optical Bloch Equations (OBE) indicates that the lineshape asymmetry is a result of fast decoherence in the system and the narrow resonances originate from coherent population pulsations that are induced by decoherence in the system. Using the OBE, we estimate that the decoherence time of the optically induced dipole (formed between the unexcited ground state the excited electron-hole pair) at room temperature is 125 fs, corresponding to a linewidth of $\sim$10 meV. The decay time of the excitation is $\sim$5-10 ns, depending on the excitation energy.  The lineshapes are well fit with the OBE indicating that the resonances are characterized by discrete levels with no evidence of many body physics.   
\end{abstract}

\maketitle


\section{\label{sec:level1}Introduction} 
Coherent nonlinear optical spectroscopy can be used to understand and characterize optical physics related to the origin of the nonlinear optical response and fundamental decay parameters in a system beyond what is measurable from more conventional measurements such as linear absorption or photoluminescence.  In the absence of many-body physics, the third order ($\chi^{(3)}$) nonlinear absorption spectrum typically shows narrow resonances arising from spectral hole burning that can be used to simultaneously understand the degree of inhomogeneous broadening, dipole decoherence rates and population decay rates\cite{Shen84,Berman}. In atomic gases, resonances in the third order nonlinear optical spectrum that are induced by decoherence from environmental interactions such as pressure-induced extra resonances have been predicted theoretically\cite{Bloembergen1978} and observed experimentally\cite{Prior1981} in high resolution four-wave mixing spectroscopy. Narrow resonances in the coherent nonlinear optical spectrum of atomic gas systems can also occur due to the effects related to collisions in the presence of quantum state specific reservoir coupling\cite{Lam1982,Steel1987}.

 Studies featuring coherent nonlinear spectroscopy techniques have been more recently focused on understanding the dynamics between optically excited excitons in solid state systems and the surrounding environment.  One of the key drivers for this research is the utilization of optically controlled exciton states in coherent control applications such as quantum information processing and all-optical switching that requires a detailed knowledge of the exciton intrinsic decay rates. Coherent optical control of excitons was demonstrated in single III-As quantum dots\cite{Chen2000} following detailed characterization of the coherent nonlinear optical response\cite{Bonadeo1998}. Spectral hole burning spectroscopy has been used to characterize the intrinsic decay rates of excitons in ensembles of nanocrystals\cite{Palinginis2003} and to observe narrow population pulsations in 2-D material\cite{Schaibley2015}. High resolution nonlinear optical techniques have also been used to probe effects such as exciton diffusion and migration \cite{Lawson1981,Wang1990}. Theoretical studies have predicted that narrow resonances in the coherent nonlinear absorption spectrum of solid state systems can occur due to exciton decoherence\cite{Loring1986}, similar to those observed in atomic gases. 
 
III-nitride systems are well known for solid state lighting and laser applications, however the high temperature stability of excitons in the system has recently thrust III-nitride nanostructures into research towards room temperature quantum photonics applications\cite{Holmes2014,Deshpande2014} where it would be required to be able to optically excite electronic states in the system. The InGaN disk-in-GaN nanowire (DINWs) system has emerged as a good potential candidate for future quantum photonics applications as they can be grown free of extended defects in the active region \cite{Le2016,Nguyen2011,Hersee2011}, have a reduced internal electric field compared to other InGaN structures\cite{Zhang2014} and have demonstrated electrically injected engineered single photon emission up to 90 K\cite{Deshpande2013,Zhang2013}. Most reports related to optical physics in DINWs thus far have been at cryogenic temperature\cite{Deshpande2013,Zhang2013,Deshpande2015}. The potential for future quantum photonics applications at room temperature in DINW systems will require a detailed understanding of the physics and intrinsic decay rates of optically excited electron-hole pairs at room temperature, properties that are relatively unknown in this system at this point. 

In this paper, we report on the high resolution (sub-$\mu$eV) third order ($\chi^{(3)}$) nonlinear optical absorption spectrum for an ensemble of InGaN disks in GaN nanowires (DINWs) at room temperature. From this spectrum, we can extract important new physics\cite{Berman,Bloembergen1978,Prior1981,Lam1982,Steel1987, Bonadeo1998,Schaibley2015,Xu2007} as mentioned earlier. The disks-in-nanowires show an inhomogeneously broadened nonlinear optical absorption spectrum ($\sim$100 meV width), measured using modulated absorption techniques, which is expected to be broadened due to variations in the morphological properties of the disks. The high resolution optical measurements, performed at energies within the modulated absorption distribution, reveal narrow sub-$\mu$eV resonances that are attributed to coherent population pulsations. The population pulsations exhibit asymmetric line shapes, where it is found from the Optical Bloch Equations (OBE) that the asymmetry is induced by fast ($\sim$100 fs) decoherence in the system.  The resonances are designated “extra” in the original analysis because they are not present in the nonlinear optical spectrum in the absence of decoherence beyond population decay because of interference of competing terms in the third order polarization that gives rise to the scattered electromagnetic field\cite{Bloembergen1978,Prior1981}.

\section{Results}

For the high resolution nonlinear optical measurements, two frequency stabilized c.-w. lasers with a mutual coherence bandwidth $\sim$2-3 MHz are focused onto the sample surface and the nonlinear signal is detected in transmission, using phase-sensitive detection as described below.  One of the beams is fixed in energy (labeled $\omega_{1}$) while the other beam (labeled $\omega_{2}$) is detected in the far field and scanned over resonances excited by the $\omega_{1}$ beam. The signal of interest is proportional to the imaginary part of the polarization that is determined by the third order optical susceptibility $\chi_{sijk}^{(3)}$($\omega_{s}$ = $\omega_{i}$ - $\omega_{j}$ + $\omega_{k}$)$E_{i}E_{j}^{*}E_{k}$, where $\omega_{s}$ = $\omega_{2}$ = $\omega_{2}$ - $\omega_{1}$ + $\omega_{1}$ and ($k_{s}$ = $k_{2}$) so that the signal is emitted along the direction of the scanning beam $\omega_{2}$.  The nonlinear signal is homodyne detected with the transmitted $\omega_{2}$ beam on a photodiode, so that the photocurrent signal of interest is proportional to $\chi_{2211}^{(3)}$($\omega_{2}$)$|E_{1}|^{2}|E_{2}|^{2}$. To ensure that the signal is in the $\chi^{(3)}$ limit and no higher order terms (such as 5th order nonlinearities, etc.) are significant, the intensities of the $\omega_{1}$ and $\omega_{2}$ beams are kept sufficiently low so that the signal scales linearly with the $\omega_{1}$ and $\omega_{2}$ intensities and measured parameters do not change with intensity.

Nonlinear signals are detected using phase-sensitive differential transmission (dT/T). The $\omega_{1}$ and $\omega_{2}$ beams are sent through two acousto-optic modulators that are modulated by two separate function generators at frequencies $\Omega_{1}$ and $\Omega_{2}$. The third order nonlinear signal is detected at the difference of the modulation frequencies $|\Omega_{1} - \Omega_{2}|$ using a lock-in amplifier.

The sample under study in this work is an ensemble of selective area InGaN disks in GaN nanowires (DINWs) grown using molecular beam epitaxy (MBE) techniques. A growth mask of hole-patterned titanium film layer was used on an GaN/sapphire substrate in order to fabricate the GaN nanowires on desired site with uniform diameter. The growth conditions for the GaN nanowire were used as a substrate temperature of 830 $^{\circ}$C with a gallium (Ga) beam equivalent pressure (BEP) of 2.1$\times$ $10^{-7}$ Torr and a nitrogen flow rate of 0.55 standard cubic centimeter per minute (sccm). The substrate temperature for InGaN disk/GaN barrier regions was reduced to 640 $^{\circ}$C with a Ga BEP of 1.5 $\times 10^{-8}$ Torr and In BEP of 1.6 $\times 10^{−7}$ Torr, respectively. Further details of the growth procedure can be found in Refs. \cite{Ra2016,Albert2013}. The InGaN disks are 3 nm thickand are grown on $<0001>$ direction along the polar c axis of GaN structure. Each GaN nanowire contains 8 InGaN disks that are separated by 3 nm of GaN barrier. The GaN nanowire is tapered at the end as part of the growth procedure so that several of the InGaN disks within each nanowire have varying diameters.

In this work, modulated absorption spectroscopy is based on exciting electron-hole pairs into the continuum energy levels of the InGaN active region. This is achieved by replacing $\omega_{1}$ with a modulated c.-w. blue (3.06 eV) non-scanning laser while probing the nonlinear absorption using a resonant $\omega_{2}$ ($\sim$2.0 eV) scanning laser again using phase-sensitive detection\cite{Pollak1993,Bonadeo1999}.  Again, the intensities of $\omega_{1}$ and $\omega_{2}$ are kept sufficiently low so that the nonlinear signal can be described as a third order signal ($\chi^{(3)}$). Within the field of view of the measurement ($\sim$1.5$\mu m^{2}$ focal spot), $\sim$200-300 DINWs are optically excited.

\begin{figure}[h]
\includegraphics[scale=0.5]{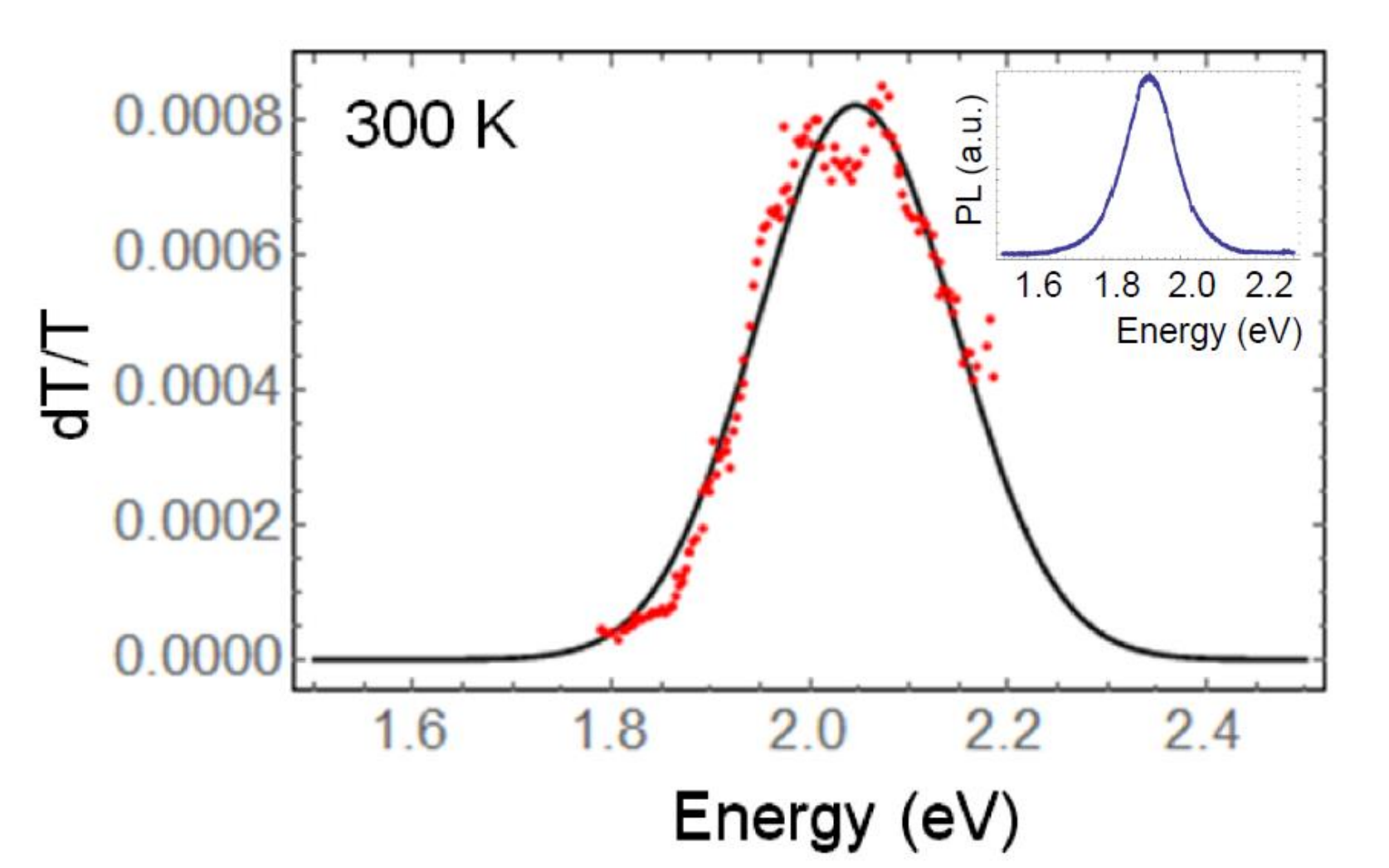}
\caption{\label{fig:epsart} Modulated absorption data (red points) fit to a Gaussian (black line) at 300 K. The Gaussian has a FWHM of $\sim$140 meV. Inset:  Photoluminescence at 300 K. }
\end{figure}

The modulated absorption data is shown in Fig. 1. The data (red points) shows a broad nonlinear absorption spectrum that is composed of several resonances. The resonances are distributed around a peak near 2.05 eV. As an approximation, the distribution is fit to a Gaussian (black line) that has a width of $\sim$140 meV. Inhomogeneous broadening in the system is expected\cite{Zhang2013,Sekiguchi2010} and can be caused by slight variations in the inter-DINW thickness, diameter or InN concentration. Within each nanowire, the tapered design can give additional inhomogeneous broadening because the energies of transitions within the DINWs along the tapered end are heavily affected by the DINW diameter. Generally, transitions for smaller diameter DINWs are blue-shifted compared to larger diameter DINWs due to strain relief effects\cite{Zhang2014}. The sample photoluminescence (PL) is shown in the inset of Fig. 1. The linewidth of the PL is similar to the modulated absorption, however the peak of the photoluminescence is red-shifted from the peak of the modulated absorption by $\sim$100 meV. The redshift of the sample PL peak from the nonlinear absorption has been observed in previous studies and is likely related to sample disorder\cite{Kundys2006}. 

\begin{figure}[h!]
\includegraphics[scale=0.45]{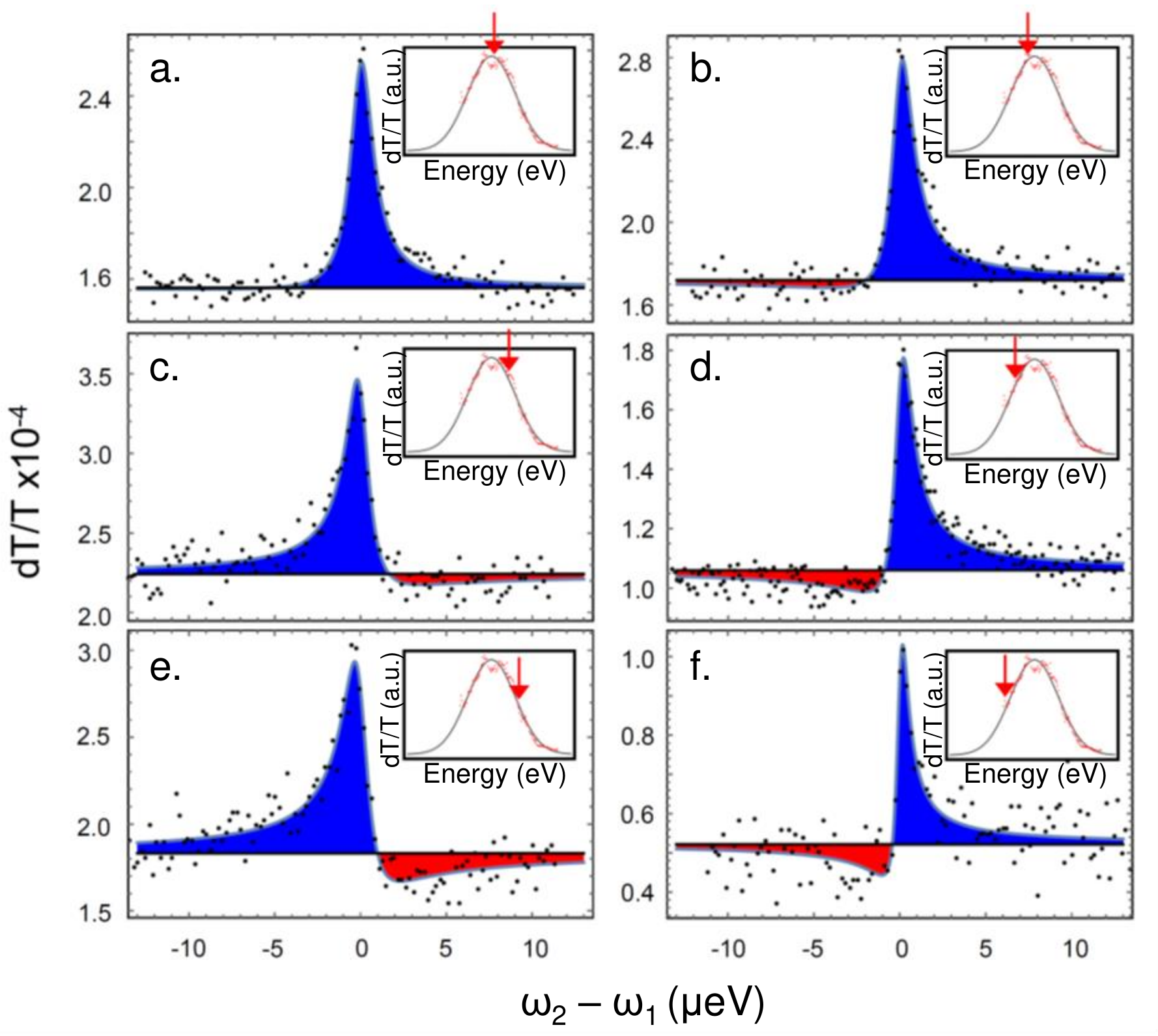}
\caption{\label{fig:fig2} Third order ($\chi^{(3)}$) coherent nonlinear absorption spectrum at 300 K. The fixed energy of $\omega_{1}$ is shown for each panel by the red arrow. The modulated absorption spectrum from Fig. 1 is shown in the inset of each panel and the $\omega_{1}$ energy corresponding to the nondegenerate nonlinear absorption spectrum is shown in each panel, within the modulated absorption spectrum, represented by the red arrow. The solid lines are a least squared fit of the imaginary part of Eq 1.}
\end{figure}

The high resolution nonlinear absorption spectrum for co-polarized $\omega_{1}$ and $\omega_{2}$ fields at room temperature is shown in Fig. 2 for several different $\omega_{1}$ energies (red arrows) that are within the energies spanned by the modulated absorption distribution. The modulated absorption distribution from Fig. 1 is shown in the inset of each panel and the arrow in each inset gives the energy of $\omega_{1}$ within the modulated absorption distribution.  In each panel, a narrow resonance is observed as well as a positive dT/T offset that corresponds to a broad resonance (having a width similar to the modulated absorption distribution) underneath each narrow resonance. When the wavelength of the $\omega_{1}$ optical field is tuned near the center of the modulated absorption distribution (top panels in Fig. 2), the narrow resonance is reasonably well described by single Lorentzian line. When the wavelength of the $\omega_{1}$ optical field is tuned above or below line center, a clear deviation (evident even in Fig. 2a and 2b) from a Lorentzian emerges:  the narrow resonance acquires a clear asymmetry where the sign of the asymmetry depends on whether $\omega_{1}$ is tuned to the high energy or low energy side of the modulated absorption distribution. The asymmetry becomes more pronounced as a function of increased detuning from line center of the modulated absorption distribution. 

\section{Discussion}

The nonlinear optical response of excitons in bulk direct bandgap semiconductors just below the band edge is known to be dominated by complex many body 
physics and cannot be interpreted or predicted by application of the OBE for two or few level atomic systems\cite{Hu1994}.  Unlike in bulk, semiconductor nanostructures such as quantum dots and 2-D materials have shown that, with sufficient quantum confinement, many body physics is suppressed and nanostructures have the potential to exhibit the features fully consistent with the OBE\cite{Bonadeo1998, Schaibley2015}.  

To qualitatively understand the physical origin of the behavior in Fig. \ref{fig:fig2}, the DINW ensemble optical response is modeled using the OBE for an inhomogeneously broadened resonant two-level system.  We will show that, based on the agreement between our data and the OBE theory, the OBE provides a good description of the optical dynamics in the DINWs, which is likely due to a large confinement energy in the system. For analytical simplicity, we consider a model in which the inhomogeneous distribution of transition frequencies $\omega$ is assumed to be a Gaussian given by $W(\omega)=  1/(\sqrt{\pi} \sigma_{W} ) e^{(-[(\omega-\omega_{0})/\sigma_{W} ]^{2} )}$. Each two-level system that composes the inhomogeneous distribution of transition frequencies represents a single DINW where the ground state  is the crystal ground state of the DINW and the excited state  represents the creation of an electron-hole pair.  The decay rate of the off-diagonal density matrix elements (i.e., the optically induced polarization or coherence) is taken to be $\gamma$ = $\gamma_{2}$/2 + $\Gamma_{dec}$, where $\Gamma_{dec}$ is the rate of pure dephasing (i.e., the result of phase changing due to the coupling to the reservoir that is separate from the decay of specific eigenstates in the systems, such as $\gamma_{2}$) and $\gamma_{2}$ is the excited population decay rate.  Using a perturbative solution to the OBE for weak optical fields in the rotating wave approximation and assuming a scaler form for the optical coupling, an expression for the third order off-diagonal density matrix element generated by two laser fields in a Gaussian distribution of homogeneously broadened two-level systems is given by \cite{Berman,Wang1991}: 
	
\begin{widetext}
\begin{eqnarray}
{\rho_{21}^{(3)}(\omega_{1},\omega_{2})}=&&-2i(\frac{\mu}{2\hbar})^{3}e^{i(\bf{k}\cdot\bf{x} - \omega_{2} t)}\frac{\sqrt{\pi}}{\sigma_{W}}
 [
   \frac{1 + \frac{2\Gamma_{dec}}{\gamma_{2}}}{\omega_{2} - \omega_{1} + 2i\gamma}(\frac{1}{\omega_{2} - \omega_{1}}[A(z_{1}) + A^{*}(z_{2})] + \frac{1}{2i\gamma}[A(z_{1}) + A^{*}(z_{1})]) \nonumber\\
   &&(1 - \frac{2i\Gamma_{dec}}{\omega_{2} - \omega{1} -i\gamma_{2}})([\frac{1}{\omega_{2} - \omega{1} + 2i\gamma}]^{2}[A(z_{1}) + A^{*}(z_{2})] - \frac{1}{\sigma_{W}(\omega_{2} - \omega_{1} + 2i\gamma)}\frac{\partial A(z_{1})}{\partial z_{1}})],
\end{eqnarray}
\end{widetext}

where $\mu$ is the transition moment, $A(z)$ = $\frac{i}{\pi}\int_{-\infty}^{\infty}\frac{e^{-x^{2}}}{z \pm x}dx$ with $Im(z)$ $>$ 0, $z_{1}$ = -$\frac{\omega_{1} - \omega_{0} -i\gamma}{\sigma_{W}}$ and $z_{2}$ = -$\frac{\omega_{2} - \omega_{0} -i\gamma}{\sigma_{W}}$.  The nonlinear polarization is then given by P = Tr[$\hat{\rho} \bf{\mu}$] (where $\bf{\mu}$ is the dipole moment operator, $\hat{\rho}$ is the density matrix operator), and the signal of interest is proportional to the imaginary part of Eq. 1.

In Fig. 2, the solid black curve is a least squares fit of the imaginary part of Eq. 1. To understand the physics behind the model, we plot the theoretical description in Eq.1 for a number of different parameters in Fig. 3. We take the width of the Gaussian distribution to be $\sigma_{W}$ = $10^{5}\gamma_{2}$ and plot the nonlinear response as a function of the detuning ($\omega_{2}$ - $\omega_{1}$) in units of $\gamma_{2}$ for several different values of the pure decoherence rate $\Gamma_{dec}$. in Fig. 3a and different $\omega_{1}$ energies for a fixed value of $\Gamma_{dec}$ in Figs. 3b. In Fig. 3a, the $\omega_{1}$ energy is fixed near the wings of the inhomogeneous Gaussian distribution at a detuning $\omega_{1}$ - $\omega_{0}$ = 1.8$\sigma_{W}$.  In the sample we cannot experimentally control the decoherence in this system, but for purposes of clarity, we show the theory for different decoherence rates. Note that a positive dT/T signal corresponds to a decrease in absorption of the $\omega_{2}$ field due to the presence of $\omega_{1}$. 
	Equation 1 contains two resonances that contribute to the nonlinear response. The population pulsation resonance is the second bracketed term on the right-hand side of Eq, 1 and that is multiplied by a factor of $[1+(2i\Gamma_{dec})/( (\omega_{2}-\omega_{1} )-i\gamma_{2})]$ and originates from interference between the $\omega_{1}$ and $\omega_{2}$ beams in first and second order of perturbation theory.  The first bracketed terms on the right-hand side of Eq. 1 describes a resonance associated with incoherent spectral hole burning. 

\begin{figure}[h]
\includegraphics[scale=0.5]{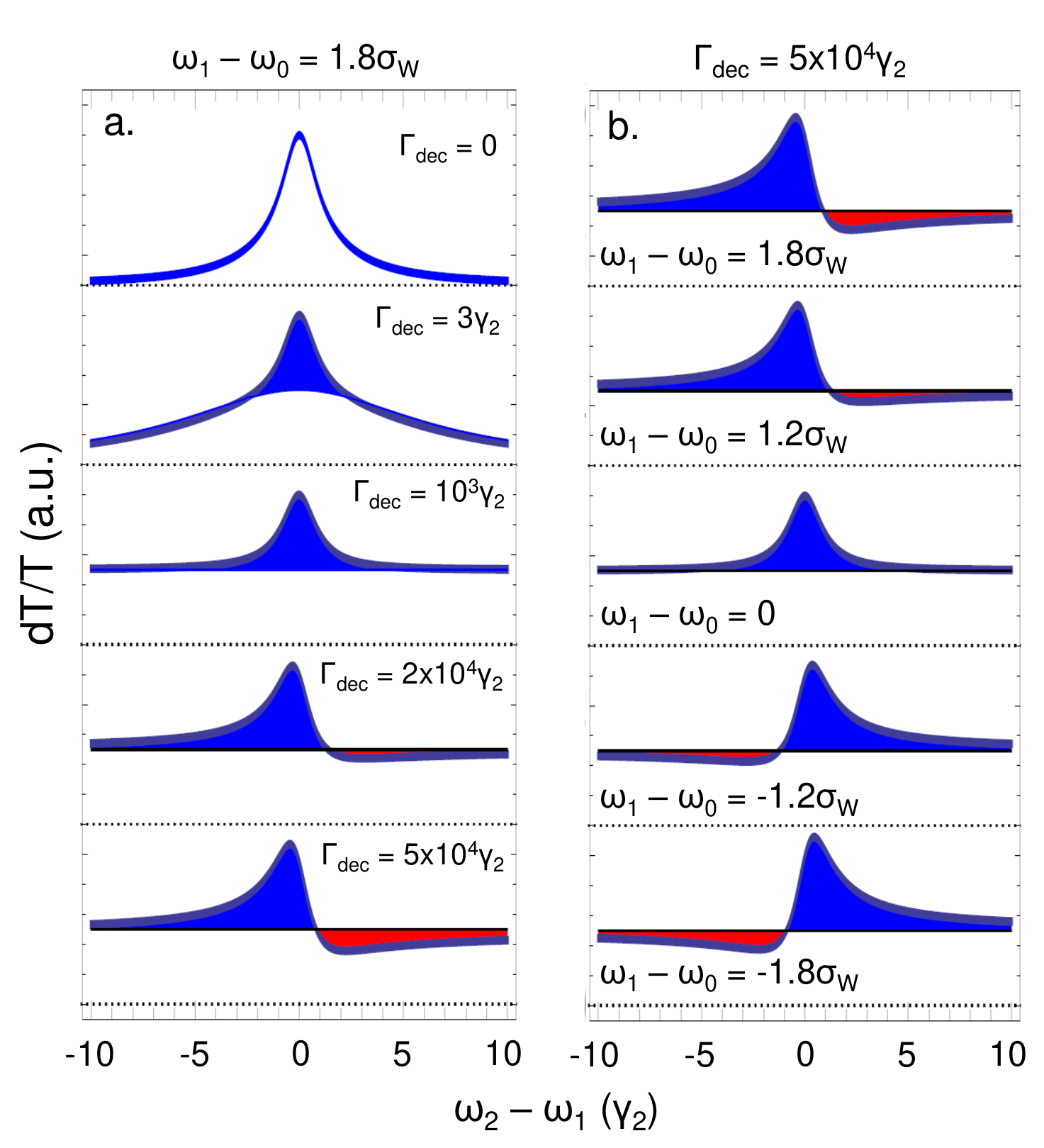}
\caption{\label{fig:fig3} Normalized theoretical differential transmission signal for an inhomogeneously broadened ensemble of two-level systems (assuming a Gaussian distribution with inhomogeneous bandwidth $1\times10^{5}\gamma_{2}$), proportional to the part of Eq. 1 for a fixed $\omega_{1}$ energy as a function $\omega_{2}$ in units of $\gamma_{2}$. $\bf{a.}$ dT/T plotted as a function of decoherence rate Γdec for a fixed detuning ($\omega_{1}$ – $\omega_{0}$)  = 1.8$\sigma_{W}$. $\bf{b.}$ dT/T plotted as a function of $\omega_{1}$ detuning for a fixed decoherence rate $\Gamma_{dec}$ = $5\times10^{4}\gamma_{2}$. The dotted black lines in each plot show zero signal. The color shows the relative contributions of the population pulsation term and the hole burning term so that the solid blue region represents the contribution of the population pulsation term relative to the hole burning term (red is a lower relative signal).}
\end{figure}

Fig. \ref{fig:fig3}a is colored to show the relative contributions of the population pulsation term and the hole burning term so that the solid blue region represents the contribution of the population pulsation term relative to the hole burning term (red is a lower relative signal). When $\Gamma_{dec}$ = 0 and $\sigma_{W}$ $>>$ $\gamma_{2}$ the total nonlinear response can be described by a single Lorentzian with a width given by twice the transition linewidth 2$\gamma$= $\gamma_{2}$. When $\Gamma_{dec}$ $>$ 0, the nonlinear absorption shows both the hole burning component (broad resonance with FWHM $\sim\gamma$) and a narrow resonance (FWHM  $\sim\gamma_{2}$) that is on top of the hole burning resonance.  The narrow resonance comes from the term $(2〖i\Gamma_{dec})/( (\omega_{2}-\omega_{1} )-i\gamma_{2})$ that originates from the population pulsation component and is a decoherence-induced resonance (not present in the absence of pure decoherence).  Seen in the density matrix picture, the physics associated with decoherence-induced extra resonances has to do with the non-cancellation of terms contributing to the nonlinear optical polarization and hence the signal\cite{Bloembergen1978,Prior1981,Khitrova1988,Rothberg1987}.  However, a deeper physical understanding emerges from the full quantum treatment where it is seen that the signal arises from contributions of non-energy conserving terms allowed by the presence of decoherence\cite{Grynberg1989,Berman1989}. A review of decoherence induced effects in nonlinear spectroscopy is reviewed in Ref. \cite{Rothberg1987}. As the rate of decoherence increases, so that $\sigma_{W}$ $>>$ $\Gamma_{dec}$ $>>$ $\gamma_{2}$ the hole burning resonance becomes essentially a constant dT/T offset compared to the population pulsation component that is dominated by the $(2〖i\Gamma_{dec})/( (\omega_{2}-\omega_{1} )-i\gamma_{2})$ resonance.  When $\Gamma_{dec}$ becomes comparable to $\sigma_{W}$ $>>$ $\gamma_{2}$, the population pulsation resonance (on top of the very broad hole burning resonance that can be treated essentially as a constant) shows an asymmetric line shape and becomes more asymmetric as the rate of pure decoherence increases.

In Figs. 3b the nonlinear absorption in the fast decoherence limit is shown for several different $\omega_{1}$ energies tuned within the inhomogeneous distribution.  The asymmetry of the population pulsation resonance exhibits a sign change depending on whether $\omega_{1}$ is tuned to above or below line center of the linear absorption and becomes more asymmetric as the magnitude of the detuning increases so that near the center of the inhomogeneous distribution the asymmetry disappears. To understand the origin of the energy-dependent lineshape asymmetry, we consider the nonlinear response in the limit that $\Gamma_{dec}$ $>$ $\sigma_{W}$ $>>$ $\gamma_{2}$. In this case, the two-level system nonlinear response can be approximately described by the third order off-diagonal density matrix element for a homogeneously broadened two-level system \cite{Berman}:
\begin{widetext}
\begin{eqnarray}
{\rho_{21}^{(3)}(\omega_{1},\omega_{2})} =-2i(\frac{\mu}{2\hbar})^{3}e^{i(\bf{k}\cdot\bf{x} - \omega_{2} t)}\frac{\sqrt{\pi}}{\sigma_{W}} \nonumber\\
\times[\frac{1}{(\omega_{0} - \omega_{2} - i\gamma)([\omega_{0} - \omega_{1}]^{2} + \gamma^{2})}(1 + \frac{2\Gamma_{dec}}{\gamma_{2}}) + \frac{1}{(\omega_{0} - \omega_{2} - i\gamma)^{2}(\omega_{0} - \omega_{1} + i\gamma)}(1 - \frac{2i\Gamma_{dec}}{\omega_{2} - \omega_{1} - i\gamma_{2}})]
\end{eqnarray}
\end{widetext}
The population pulsation term of interest is again given by the imaginary component of the second term on the right hand side of Eq. 2. 

For small values of detuning so that $\omega_{1}$, $\omega_{2}$ $\approx$ $\omega_{0}$ ($\Gamma_{dec}$ $>>$ $\gamma_{2}$), Eq. 2 is approximately given by  $\frac{2i}{\gamma^{3}}(\frac{2i\Gamma_{dec}}{\omega_{2} - \omega_{1} -i\gamma_{2}})$.  The imaginary part gives a Lorentzian with a FWHM of 2$\gamma_{2}$. When $|\omega_{1}$ - $\omega_{0}|$ and $|\omega_{2}$ - $\omega_{0}|$ $\sim$ $\gamma$, the line shape becomes a weighted combination of the narrow resonance Lorentzian and the corresponding dispersive part, hence the asymmetry. The wavelength-dependent lineshape asymmetry discussed above is then a signature of decoherence-induced population pulsation resonances in two-level systems and is not observed in the saturation term (the first term in Eq. 1). 

Based on a comparison of the $\omega_{1}$ wavelength dependence of the theoretical population pulsation line shapes in Fig. 3 to the data in Fig. 2, we assign the narrow resonances in Fig. 2 to coherent population pulsations. The population pulsation resonances are decoherence-induced extra resonances as noted above, where in this case the theory from equation 1 and Fig. 3 indicates the rate of decoherence $\Gamma_{dec}$ is comparable to the inhomogeneous bandwidth from Fig. 1, thus giving an asymmetric dT/T line shape.  Measurement of the asymmetric resonance allows us to simultaneously measure the wavelength-dependent rate of decoherence and the wavelength-dependent population decay rate $\gamma_{2}$, a measurement that is normally more difficult using photoluminescence techniques. The population decay rate for an excitonic system is given by the sum of the radiative and non-radiative decay rates of the optically excited excitons back to the crystal ground state. The decay times $\gamma_{2}^{-1}$ extracted from the fits in Fig. 2 are $\sim$5-10 ns and are dependent on excitation energy. The energy dependence of $\gamma_{2}$ will be discussed below. The values of the decay time obtained in this study are consistent with decay times reported in similar samples using time-resolved photoluminescence techniques\cite{Zhang2013,Deshpande2015}.

The decoherence rate $\Gamma_{dec}$ of optically excited electron-hole pairs is estimated from an average of the fit parameters from Fig. 2 to be $\Gamma_{dec}^{-1}$ $\sim$125 fs, with significant variation ($\pm$ 38 fs) as a function of the frequency of $\omega_{1}$, perhaps reflecting the sample inhomogeneity. In principle, the value of the inhomogeneous bandwidth $\sigma_{W}$ could be a fixed constant for each fit to the data based on the measurements in Fig. 1.  The average inhomogeneous bandwidth extracted from the fits ($\sim$75 $\pm$ 12 meV) is smaller than the bandwidth expected from the Gaussian fit to the modulated absorption data in Fig. 1. We attribute this discrepancy to the fact that the inhomogeneously broadened nonlinear absorption signal in Fig. 1 is more complex than a single Gaussian as seen by a close inspection of the PLE data in Fig. 1. A more accurate description of inhomogeneous distribution would likely involve using a kernel function in Eq. 1 with multiple resonances, where each would have a linewidth less than the Gaussian fit in Fig. 1. We did not observe a clear correlation between the ratio of the decoherence rate to the inhomogeneous bandwidth $\frac{\Gamma_{dec}}{\sigma_{W}}$ as a function of $\omega_{1}$.  Ultrafast decoherence ($\Gamma_{dec}^{-1}$ $\sim$100 fs) has been reported in time domain studies of optically excited states in InGaN/GaN quantum well systems at low temperature\cite{Kundys2006} that was attributed to effects related to material disorder\cite{Lonsky1989}. At room temperature, effects due to the interactions of optically excited electron-hole pairs with acoustic and optical phonons can create an additional source of decoherence\cite{Krummheuer2002}. The temperature dependence of the decoherence rate of optically excited electron hole pairs in the DINW system will be the subject of future studies. 

Figure 4 shows a plot of the dependence of the decay rate $\gamma_{2}$ as a function of $\omega_{1}$ (in energy units) superimposed on the modulated absorption data.  On the low energy side of the modulated absorption distribution the decay rate decreases as a function of decreasing energy, however it is relatively independent of energy on the high energy side of the modulated absorption distribution. Similar energy-dependent lifetime effects have been reported in a bulk\cite{Cohen1982} and multi-quantum wells\cite{Hegarty1984}, where the effect was attributed to the existence of a “mobility edge” that occurs at the line center of the exciton transition. Effects such as coupling of optically excited electron-hole pairs to charge traps or other states that can provide additional decay channels (where the rate of decay is $\sim\gamma_{2}$) in the system that would give additional narrow population pulsation resonances\cite{Schaibley2015}. These effects do not appear to be significant in this sample as the population pulsations in Fig. 2 fit well to a single resonance. Hence, an understanding of this behavior remains unclear and requires further investigation.
\begin{figure}[hbt]
\includegraphics[scale=0.35]{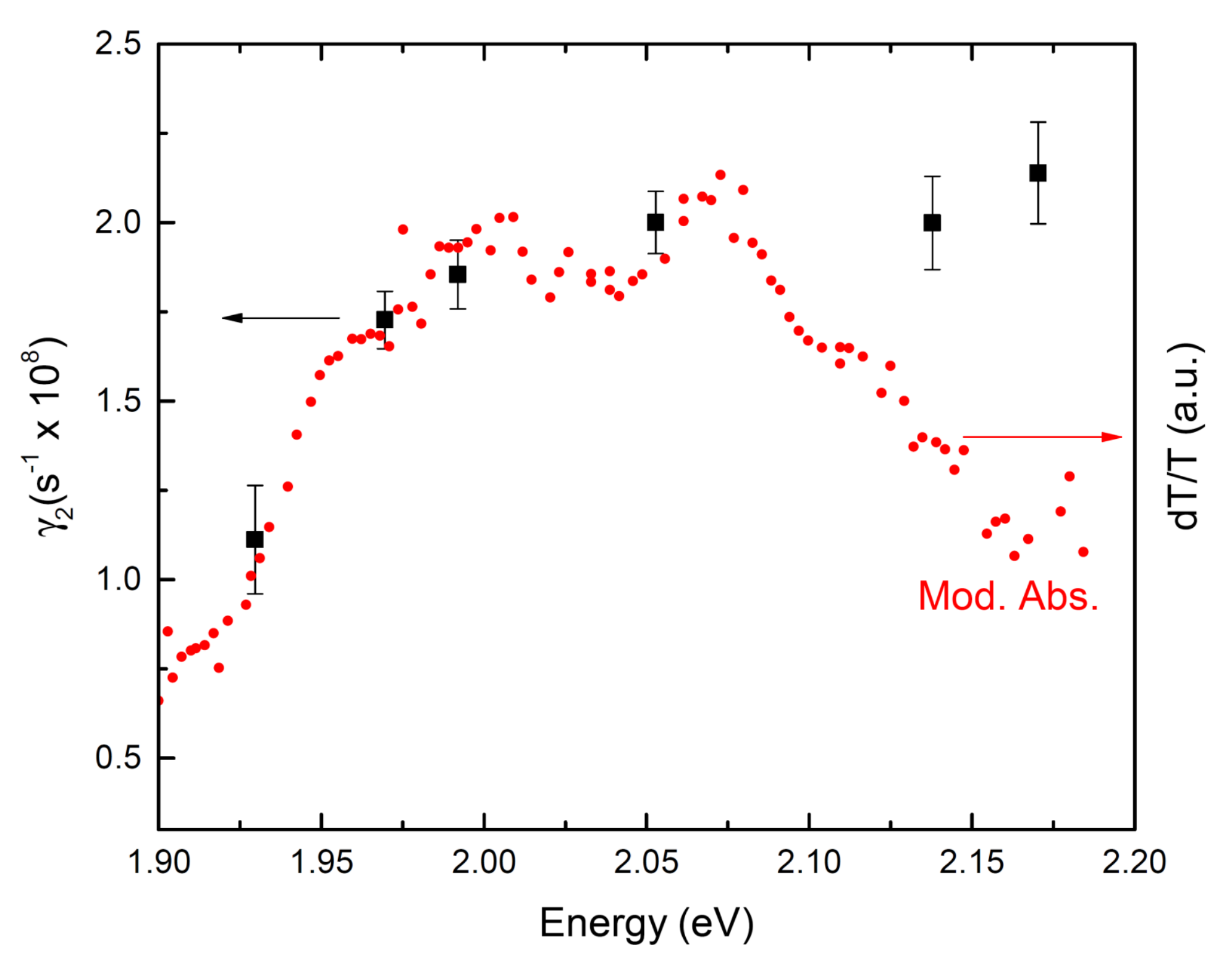}
\caption{\label{fig:fig4} Population decay rates $\gamma_{2}$ as a function of $\omega_{1}$ energy obtained from the fits of equation 1 to the data in Fig. 2 and the modulated absorption distribution from Fig. 1. The uncertainties in the population decay times are from the error to the fit.}
\end{figure}

\section{Conclusion}

In summary, we have observed narrow (sub-$\mu$eV) resonances in the high resolution nonlinear absorption spectrum of an ensemble of InGaN DINWs that are attributed to coherent population pulsations. The population pulsation resonances, measured at energies within the inhomogeneously broadened nonlinear absorption spectrum, show asymmetric line shapes that become more pronounced as a function of detuning from line center of the inhomogeneous distribution. The asymmetric population pulsations are assigned to decoherence-induced extra resonances based on a qualitative comparison of the data to the nonlinear response expected from a Gaussian inhomogeneous distribution. The qualitative agreement with the OBE is evidence that the DINWs suppress many body physics in the nonlinear response.  We estimate that the inverse decoherence rate of optically excited electron-hole pairs is $\sim$125 fsec at room temperature, comparable to the inhomogeneous bandwidth. The energy-dependent exciton population decay rates $\gamma_{2}$ extracted from the population pulsation measurements suggests the possible existence of a mobility edge in the system that occurs near the center of the modulated absorption distribution. This study lays the groundwork for future research towards room temperature quantum photonics applications in InGaN such as quantum information processing or all-optical switching as the quality of the material improves.

\begin{acknowledgments}
The work was supported in part by the National Science Foundation (NSF) through CPHOM (DMR 1120923).  
\end{acknowledgments}

\bibliography{PRB_bib}

\end{document}